\documentclass[preprint,12pt]{elsarticle}

\usepackage{color}

\usepackage{amssymb}

\newcommand{\be}{\begin{equation}}
\newcommand{\ee}{\end{equation}}
\newcommand{\bea}{\begin{eqnarray}}
\newcommand{\eea}{\end{eqnarray}}
\newcommand\pp{\,\,\,.}
\newcommand\vv{\,\,\,,}

\begin{document}

\begin{frontmatter}



\title{Updated BBN bounds on the cosmological lepton asymmetry for non-zero $\theta_{13}$}


\author[INFN]{Gianpiero Mangano}
\author[INFN,UNap]{Gennaro Miele}
\author[IFIC]{Sergio Pastor}
\author[INFN,UNap]{Ofelia Pisanti}
\author[UNap,MPI]{Srdjan Sarikas}

\address[INFN] {Istituto Nazionale di Fisica
 Nucleare - Sezione di Napoli \\ Complesso Universitario di Monte
 S.~Angelo, I-80126 Napoli, Italy}
\address[UNap] {Dipartimento di Scienze Fisiche, Universit\`{a} di Napoli
{\it Federico II} \\ Complesso Universitario di Monte S.~Angelo, I-80126
Napoli, Italy} 
\address[IFIC] {Instituto de F\'{\i}sica Corpuscular
(CSIC-Universitat de Val\`{e}ncia),\\ Ed.\ Institutos de
Investigaci\'{o}n, Apdo.\ correos 22085, E-46071 Valencia, Spain}
\address[MPI] {Max-Planck-Institut f\"ur Physik (Werner-Heisenberg-Institut)\\
F\"ohringer Ring 6, 80802 M\"unchen, Germany}


\begin{abstract}
We discuss the bounds on the cosmological lepton number from Big Bang Nucleosynthesis (BBN), in light of recent evidences for a large value of the neutrino mixing angle $\theta_{13}$, $\sin^2\theta_{13} \gtrsim 0.01$ at 2-$\sigma$. The largest asymmetries for electron and $\mu$, $\tau$ neutrinos compatible with $^4$He and $^2$H primordial yields are computed versus the neutrino mass hierarchy and mixing angles. The flavour oscillation dynamics is traced till the beginning of BBN and neutrino distributions after decoupling are numerically computed. The latter contains in general, non thermal distortion due to the onset of flavour oscillations driven by solar squared mass difference in the temperature range where neutrino scatterings become inefficient to enforce thermodynamical equilibrium. Depending on the value of $\theta_{13}$, this translates into a larger value for the effective number of neutrinos, $N_{\rm{eff}}$. Upper bounds on this parameter are discussed for both neutrino mass hierarchies.  Values for $N_{\rm{eff}}$ which are large enough to be detectable by the Planck experiment are found only for the (presently disfavoured) range $\sin^2\theta_{13} \leq 0.01$.
\end{abstract}


\end{frontmatter}


\hfill{\tt IFIC/11-58, DSF-13-2011}
\section{Introduction}
\label{intro}

Nowadays flavour neutrino oscillations are well established thanks to a plethora of 
experimental results on the detection of reactor, accelerator, atmospheric and solar neutrinos. 
Two neutrino mass-squared differences and three mixing angles drive oscillations among the three active neutrinos, all of them, except one angle, known with a precision better than 25\%
\cite{Schwetz:2011qt,Fogli:2011qn,Schwetz:2011zk}.
The last parameter which is not so well known is the mixing angle $\theta_{13}$. Until very recently only an upper bound on $\sin^2\theta_{13}$ existed, while the present year we have witnessed the first indications of non-zero $\theta_{13}$ values from the analysis of global data \cite{Fogli:2011qn,Schwetz:2011zk}, especially recent 
$\nu_\mu\to\nu_e$ searches at the T2K long-baseline experiment \cite{Abe:2011sj}. 
This result opens up the possibility of measuring, in a not so distant future, the
pattern of neutrino masses (the mass hierarchy) and a possible CP violation in the leptonic 
sector \cite{Nunokawa:2007qh}, the last remaining unknowns together with the nature of neutrinos (Dirac or Majorana).

Neutrino oscillations have implications in many research areas in particle and astroparticle physics. However, the consequences 
of non-zero neutrino mixing in cosmology are not so important, despite the fact that relic neutrinos are the second most abundant particles in the Universe, with almost the same number density as photons. The reason is well known: in first approximation all neutrino flavours were produced by frequent interactions in the early hot Universe, with the same momentum spectra. Thus neutrino oscillations, although effective right after neutrino decoupling, do not modify the properties of cosmological neutrinos (except for very small effects from non-instantaneous neutrino decoupling \cite{Mangano:2005cc}). This holds for active neutrino oscillations, while the case of active-sterile oscillations does lead
to effects on cosmological observables if one or more extra neutrino species are populated (see e.g.\  \cite{Dolgov:2003sg,Kirilova:2006wh,Melchiorri:2008gq,Hamann:2011ge}).

There exists, however, one situation where active-active oscillations have an impact on cosmological neutrinos, namely when a large flavour neutrino asymmetry was previously created.
It is usually assumed that such an asymmetry, parameterized by the number
density ratios
\be
\eta_{\nu_\alpha} = \frac{n_{\nu_\alpha}-n_{\bar{\nu}_\alpha}}{n_\gamma}\vv \,\,\, \alpha=e,\mu,\tau \vv
\label{etanualpha}
\ee
should be of the same order of the cosmological baryon number 
$\eta_b=(n_b-n_{\bar{b}})/n_\gamma$, due to the equilibration by sphalerons of lepton and baryon 
asymmetries in the very early universe. Thus one does not expect
flavour neutrino asymmetries much larger than a few times $10^{-10}$,
the value of $\eta_b$ measured by present observations, such as 7-year data from the WMAP satellite and other cosmological measurements \cite{Komatsu:2010fb}. There are, however, some models where a lepton asymmetry orders
 of magnitude larger than the baryon one could survive, see e.g.\ 
 \cite{MarchRussell:1999ig,McDonald:1999in}, in the neutrino sector with an influence on fundamental physics in the early universe, such as the QCD transition \cite{Schwarz:2009ii}  or a potential relation with the cosmological
magnetic fields at large scales \cite{Semikoz:2009ye}. 

Present cosmological observations are not sensitive to a large neutrino asymmetry if
$|\eta_\nu|\lesssim 10^{-2}$. Only larger values lead to a significant enhancement of the contribution
of active neutrinos to the radiation energy density of the Universe or to changes in the production of light elements in Big Bang Nucleosynthesis (BBN). In particular, the primordial abundance of
$^4$He depends on the presence of an electron neutrino asymmetry and sets  a stringent bound on $\eta_{\nu_e}$ which does not apply to the other flavours unless neutrino oscillations
are effective before BBN, leaving a total neutrino asymmetry of order unity unconstrained
\cite{KangSteigman,Hansen:2001hi}. 

A decade ago, it was shown that flavour neutrino conversions in the early Universe are indeed suppressed by matter effects at large temperatures, and it is only at temperatures $T\lesssim 10$ MeV that oscillations set on and are large enough to achieve
strong flavor conversions before BBN
\cite{Dolgov:2002ab,Wong:2002fa,Abazajian:2002qx,Lunardini:2000fy}.
For the present measured values of neutrino mixing parameters, the degree of flavour
equilibration depends on the value of $\theta_{13}$. As discussed in \cite{Pastor:2008ti},
this parameter fixes the onset of flavour oscillations involving $\nu_e$'s, which in turn
determines whether neutrinos interact enough with electrons and positrons to 
transfer the excess of energy density due to the initial $\eta_{\nu_\alpha}$
to the electromagnetic plasma. Recently
we have found the BBN bounds on the cosmological lepton number for a range of
initial flavour neutrino asymmetries \cite{Mangano:2010ei}.

Prompted by the recent indication of non-zero values for $\theta_{13}$ and the hints
of possible extra radiation from cosmological data \cite{Komatsu:2010fb}, 
we have updated the analysis 
in \cite{Mangano:2010ei} with the aim of finding the BBN bounds on both the total
neutrino asymmetry and its maximum contribution to the radiation content of the Universe
in the whole range of $\theta_{13}$ values allowed by oscillation data, as well as considering both neutrino
mass hierarchies.

This letter is organized as follows. We introduce in Section \ref{sec:evo} the formalism of kinetic equations which rule the evolution of neutrino distributions and describe the dynamics
of neutrino asymmetries in the epoch just immediately BBN.
In Section \ref{sec:BBN} we study the BBN
constraints on lepton number that can obtained for a wide
range of initial neutrino asymmetries, with emphasis on the effects of
a mixing angle $\theta_{13}$ in the region experimentally favoured.
Finally in Section \ref{sec:conclusions} we give our concluding remarks.

\section{Neutrino evolution in the presence of lepton asymmetries}
\label{sec:evo}

Active neutrinos were produced in the very early Universe and their energy spectrum is kept in 
chemical and kinetic equilibrium by weak interactions until temperatures $T\simeq {\cal O}$(MeV), when the corresponding collision rates fall below the cosmological expansion rate. Therefore, 
if flavour neutrino asymmetries existed, at larger temperatures the neutrino distribution of 
momenta is a Fermi-Dirac spectrum parameterized by a temperature $T$ (the same of $e^+e^-$ and photons) and a well defined chemical potential $\mu_{\nu_\alpha}$ for
$\alpha= e,\mu,\tau$. Each flavor neutrino asymmetry in Eq.\ (\ref{etanualpha})
can be expressed in terms of the corresponding degeneracy parameter
$\xi_\alpha\equiv \mu_{\nu_\alpha}/T$ as
\be
\eta_{\nu_\alpha} = \frac{1}{12 \zeta(3)}
\left( \pi^2\xi_\alpha+\xi_\alpha^3 \right) \vv
\label{eta-eq}
\ee
with $\zeta(3)\simeq1.20206$. This expression is modified later by a factor $(T_{\nu_\alpha}/T_\gamma)^3$, when $e^+e^-$ pairs annihilate into photons. The corresponding contribution of 
neutrinos in equilibrium to the total energy density, 
usually parameterized as
$\rho_r/\rho_\gamma=1 + 7/8 (4/11)^{4/3} N_{\rm eff}$ after the $e^+e^-$ annihilation phase, 
is enhanced for non-zero neutrino asymmetries as follows
\be
N_{\rm{eff}} = 3+\sum_{\alpha=e,\mu,\tau}
\left[\frac{30}{7}\left( \frac{\xi_\alpha}{\pi} \right)^2 +
\frac{15}{7}\left( \frac{\xi_\alpha}{\pi} \right)^4 \right]  \pp \label{deltan}
\ee
The parameter $N_{\rm{eff}}$ is the {\it effective number of neutrinos} whose standard
value is 3 in the limit of instantaneous neutrino decoupling. 

We are interested in calculating the evolution of the active neutrino spectra from large temperatures, when they followed a Fermi-Dirac form, until the BBN epoch. This includes 
taking into account neutrino interactions among themselves and with charged leptons, as 
well as flavor oscillations, which become effective at similar temperatures.
In such a case the best way to describe neutrino distributions  is to use matrices in
flavor space $\varrho_{\bf p}$ \cite{Sigl:1993fn, McKellar:1994ja}. For three active neutrino species, 
we need $3\times3$ matrices in flavor space $\varrho_{\bf p}$ 
for each neutrino momentum ${\bf p}$, where
the diagonal elements are the usual occupation numbers 
and the off-diagonal ones encode phase information
and vanish for zero mixing. The corresponding
equations of motion (EOMs) for $\varrho_{\bf p}$ are the same as those
considered in references \cite{Pastor:2008ti,Mangano:2010ei}, where the reader
can find more details on the approximations made to solve them and related references, 
\begin{equation}
{\rm i}\,\frac{d\varrho_{\bf p}}{dt} =[{\sf\Omega}_{\bf p},\varrho_{\bf
p}]+ C[\varrho_{\bf p},\bar\varrho_{\bf p}]\,,
\label{drhodt}
\end{equation}
and similar for the antineutrino matrices $\bar\varrho_{\bf p}$. The last
term corresponds to the effect of neutrino
collisions, i.e.\ interactions with exchange of momenta, which are implemented 
as in ref.\ \cite{Pastor:2008ti}. These collision terms, proportional to the square of the Fermi constant $G_{\rm F}$, are crucial for modifying the 
neutrino distributions to achieve
equilibrium with $e^\pm$ and, indirectly, with photons. In the absence of neutrino mixing, the 
EOMs include only collision terms and preserve the flavour neutrino asymmetries $\eta_{\nu_\alpha}$.

The first term on the right-hand side of Eq.\ (\ref{drhodt}) describes flavor oscillations,
\begin{equation}
{\sf\Omega}_{\bf p}=\frac{{\sf M}^2}{2p}+
\sqrt{2}\,G_{\rm F}\left(-\frac{8p}{3 m_{\rm w}^2}\,{\sf E}
+\varrho-\bar\varrho\right)\,,
\end{equation}
where $p=|{\bf p}|$ and ${\sf M}$ is the neutrino mass matrix (opposite
sign for antineutrinos), which in the flavour basis is not diagonal and includes
the mixing parameters that characterizes the vacuum term of oscillations. 
In our calculations we have fixed both mass-squared differences and the angles $\theta_{12}$ and $\theta_{23}$ to the best-fit values in \cite{Schwetz:2011zk}. Varying these parameters within the allowed $3\sigma$ ranges does not modify our results. Instead, we will consider the whole presently allowed range of $\theta_{13}$ values, 
approximately from $0.001$ to $0.035-0.05$ at $3\sigma$ (depending on the reactor neutrino
fluxes, see \cite{Fogli:2011qn,Schwetz:2011zk}), 
adding the case of zero $\theta_{13}$ for comparison.

Matter effects are included via the term
proportional to the Fermi constant $G_{\rm F}$, the so-called
neutrino potentials. The one proportional to
$\varrho-\bar\varrho$, where $\varrho=\int \varrho_{\bf p}\,{\rm d}^3{\bf
p}/(2\pi)^3$ (and similar for antineutrinos) arises from neutrino-neutrino interactions and 
it was shown in \cite{Dolgov:2002ab, Wong:2002fa, Abazajian:2002qx} that
for the relevant values of neutrino asymmetries this matter term dominates 
but does not suppress flavour oscillations. Instead, this term
leads to synchronized oscillations of neutrinos and antineutrinos of 
different momenta \cite{Pastor:2001iu}.

The small baryon density in the early Universe implies that the refractive matter term
proportional to the difference between the charged lepton and antilepton number densities
(CP asymmetric) can be neglected compared to the CP symmetric term, proportional
to the sum of energy densities \cite{Sigl:1993fn,Notzold:1987ik}. This appears in
Eq.\ (\ref{drhodt}) with ${\sf E}$, the $3\times3$ flavor diagonal matrix of charged-lepton energy 
densities. The onset of flavour oscillations occurs at a the temperature at which the vacuum and charged-leptons background terms become equal in magnitude, i.e.\ when the
following terms
\begin{equation}
-\frac{\Delta m^2}{2p}\cos2\theta\qquad , \qquad
-\frac{8\sqrt{2}G_F p}{3m^2_W}(\rho_{l^-}+\rho_{l^+})
\label{vac_matter}
\end{equation}
are comparable for the relevant mixing parameters $\Delta m^2$ and $\theta$, and the 
charged-lepton energy density $\rho_l$ (see e.g.\ \cite{Wong:2002fa}). Tau 
leptons are too heavy to have a significant density at MeV temperatures, while
the energy density of $\mu^\pm$ is exponentially suppressed, leading to 
$\nu_\tau-\nu_\mu$ mixing driven by $\Delta m_{31}^2$ and $\theta_{23}$
at $T\simeq 15$ MeV \cite{Dolgov:2002ab}, when weak interactions are fully
effective. Thus, as in previous works, our numerical calculations start at $T=10$ MeV
with initial degeneracy parameters $\xi_x\equiv \xi_\mu=\xi_\tau$ and $\xi_e$.

\begin{figure}[t]
\begin{center}
         \includegraphics[width=0.85\textwidth,angle=0]{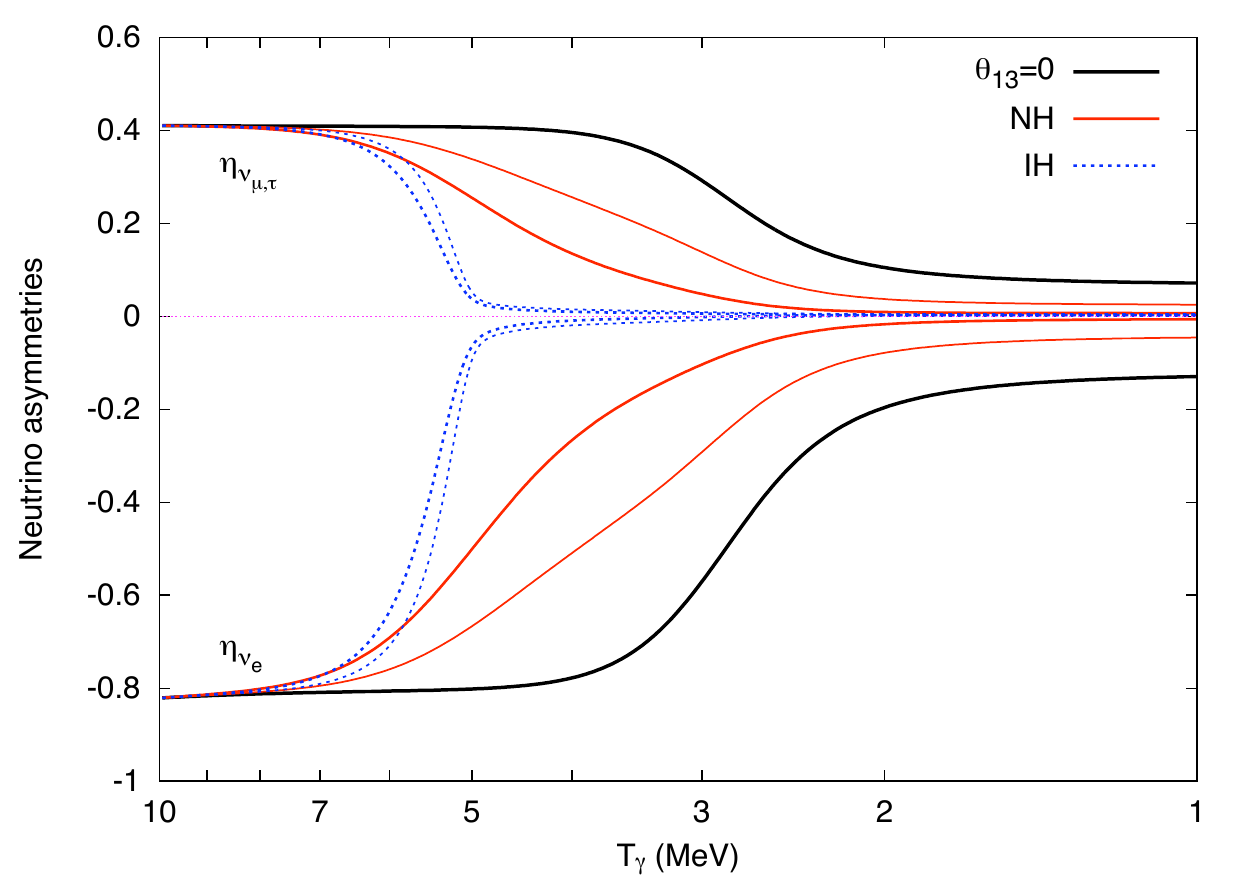}
    \end{center}
\caption{\label{evolLnu} Evolution of the flavor neutrino asymmetries when
$\eta_{\nu_e}^{\rm in}=-0.82$ and zero total asymmetry. 
The outer solid curves correspond to vanishing $\theta_{13}$ (black lines), while
the inner ones (red lines) were calculated in the NH for two values of
$\sin^2\theta_{13}$: from left to right, $0.04$ and $0.02$. The same two
values of $\sin^2\theta_{13}$ apply to the cases shown as
blue dotted lines, but in the IH.}
\end{figure}

For flavour neutrino oscillations involving $\nu_e$'s, the crucial parameters are
$\Delta m_{31}^2$ and $\theta_{13}$ because they fix the moment when the neutrino 
conversions become important, i.e.\ when the absolute values of both terms in 
Eq.\ (\ref{vac_matter}) are equal. One easily finds that this occurs at a temperature
\be
T_c \simeq 19.9 \,\left(\frac{p}{T}\right)^{-1/3}\, 
\left(\frac{|\Delta m_{31}^2|}{{\rm eV}^2}\right)^{1/6}
~{\rm MeV}
\label{Tc}
\ee
for $\cos2\theta_{13}\simeq 1$ and $e^\pm$ taken as relativistic particles. 
For $|\Delta m_{31}^2|=2.5\times10^{-3}$ eV$^2$ and an average neutrino momentum, 
one finds $T_c\simeq 5$ MeV. If $\Delta m_{31}^2>0$ (normal neutrino mass hierarchy, NH)
both terms in Eq.\ (\ref{vac_matter}) have the same sign and neutrino oscillations follow
an MSW conversion when the vacuum term overcomes the matter potential at $T\simeq T_c$.
The degree of conversion depends in this case on the value of $\theta_{13}$
\cite{Dolgov:2002ab,Wong:2002fa,Abazajian:2002qx}, being very efficient
compared with $\theta_{13}=0$ if this mixing angle presents a value close to the upper 
bound, as can 
be seen in figure \ref{evolLnu} for one particular case with zero total lepton number
and different choices of $\theta_{13}$.

The conversion for non-zero $\theta_{13}$ is more evident for the inverted mass hierarchy,
as shown in figure  \ref{evolLnu},
due to the resonant character of the MSW transition for $\Delta m^2_{31}<0$. Indeed, 
for IH the sum of the two terms in Eq.\ (\ref{vac_matter}) vanishes and the equipartition
of the total lepton asymmetry among the three neutrino flavours is quickly achieved,
even for $\sin^2\theta_{13}\lesssim 0.01$, unless of course $\theta_{13}$ is
extremely small. Finally, for negligible $\theta_{13}$ flavour oscillations are 
not effective until $T\lesssim 3$ MeV (outer lines in figure  \ref{evolLnu}), a 
value that can be found substituting  $\Delta m_{31}^2$ for
$\Delta m_{21}^2=7.6\times 10^{-5}$ eV$^2$ in
Eq.\ (\ref{vac_matter}).

The moment when flavour oscillations in the presence of neutrino asymmetries 
become effective is important not only to establish the electron neutrino asymmetry 
at the onset of BBN, but also to determine whether weak interactions 
with $e^+e^-$ can still keep neutrinos in good thermal contact with the ambient 
plasma. Oscillations redistribute the asymmetries among the flavours, but
only if they occur early enough would interactions preserve Fermi-Dirac spectra
for neutrinos, in such a way that a chemical potential $\mu_{\nu_\alpha}$
is well defined for each $\eta_{\nu_\alpha}$ and the relations in Eqs.\ (\ref{eta-eq}) 
and (\ref{deltan}) remain valid. For instance, if the initial values of the flavour asymmetries 
have opposite signs, neutrino conversions will tend to reduce the asymmetries which in turn
will decrease $N_{\rm eff}$.
But if flavour oscillations take place at temperatures
close to neutrino decoupling, this would not hold and an extra contribution of neutrinos to 
radiation is expected with respect to the value in Eq.\ (\ref{deltan}), as emphasized in
\cite{Pastor:2008ti}.

\begin{figure}[t]
\begin{center}
         \includegraphics[width=0.8\textwidth,angle=0]{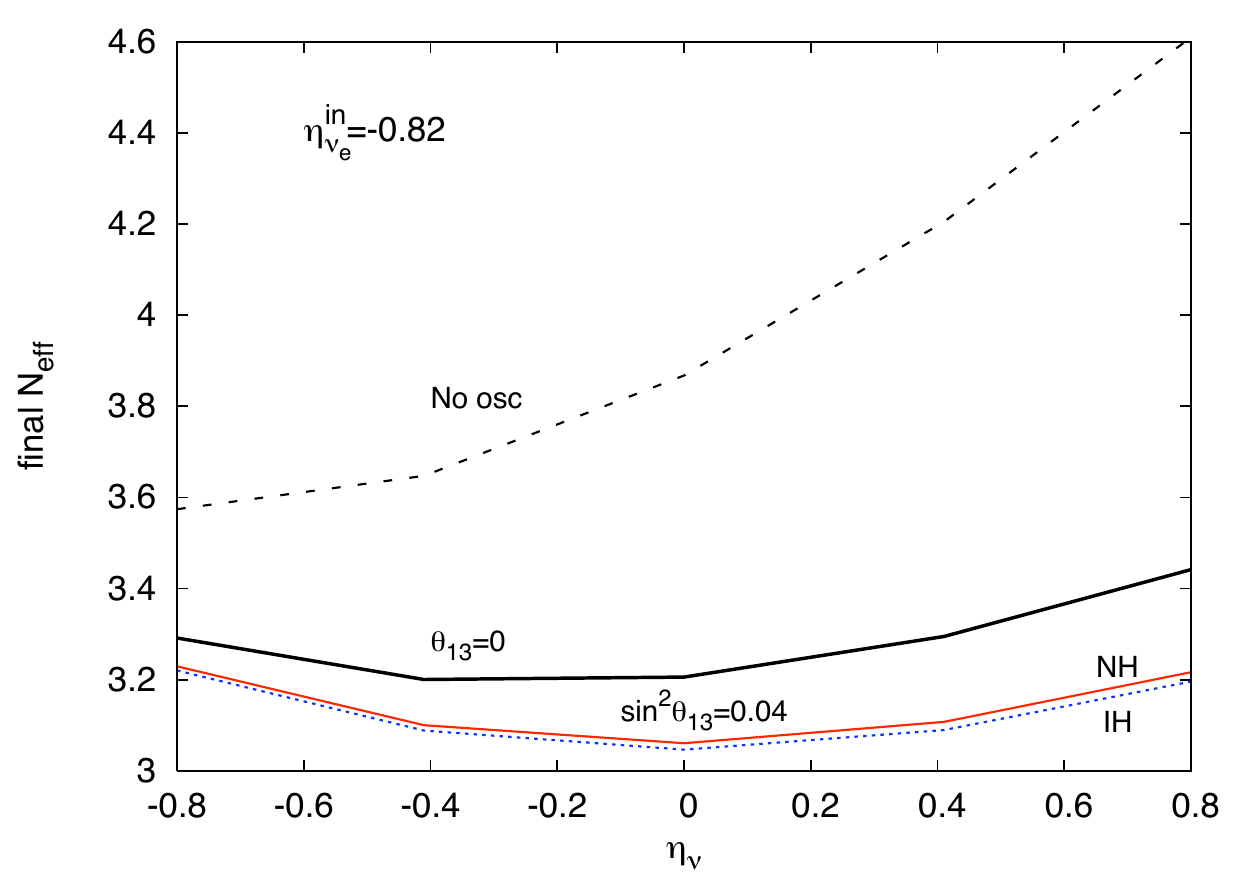}
    \end{center}
\caption{\label{finalNeff} Final contribution of neutrinos to the total radiation energy density, parametrized with $N_{\rm eff}$, as a function of the total neutrino asymmetry for a particular value of the initial electron neutrino asymmetry ($\eta_{\nu_e}^{\rm in}=-0.82$). {}From top to bottom, the various lines correspond, respectively, to the following cases: no neutrino oscillations ($\eta_{\nu_e}$ conserved), $\theta_{13}=0$, and
$\sin^2\theta_{13}=0.04$ for normal (red solid line) and inverted (blue dotted line) neutrino mass hierarchy.}
\end{figure}

A way to see the role of flavour oscillations on the reduction of the final value of $N_{\rm eff}$
from neutrino asymmetries is given in figure \ref{finalNeff}. Here we have fixed
the initial electron neutrino asymmetry to $\eta_{\nu_e}^{\rm in}=-0.82$ as in figure \ref{evolLnu},
but varied the total asymmetry in the range $-0.8\leq\eta_\nu\leq 0.8$. In the absence of
neutrino mixing the final value of $N_{\rm eff}$ is that given by Eq.\ (\ref{deltan}), directly related to the chemical potentials, and for this particular range it can be as large as $N_{\rm eff}\simeq 4.6$. 
Instead, when  oscillations are included the three flavour asymmetries are modified and the
contribution of neutrinos is largely reduced, even for $\theta_{13}=0$.
Finally, for $\sin^2\theta_{13}=0.04$ and both NH or IH, the final flavour asymmetries
are given by $\eta_{\nu_\alpha}\simeq \eta_\nu/3$. In such a case, we expect neutrinos
to almost follow Fermi-Dirac spectra and  $N_{\rm eff}$ as given in Eq.\ (\ref{deltan}). For instance,
for $\eta_\nu=0.8$ one expects $\xi_{\nu_{e,\mu,\tau}}\simeq 0.38$ and
a total contribution to the radiation energy density of $N_{\rm eff}\simeq 3.19$, very close
to what we find in our numerical calculations for $\sin^2\theta_{13}=0.04$, while
we found $N_{\rm eff}\simeq 3.43$ for $\theta_{13}=0$.

\section{BBN bounds on the cosmological lepton asymmetry}
\label{sec:BBN}

Cosmological neutrinos influence the primordial production of light elements
in two ways. First of all, they contribute to the radiation energy density that fixes the 
expansion of the Universe during BBN, a background effect that is parameterized with
$N_{\rm eff}$. For the particular case of $^4$He, its primordial abundance is enhanced 
for $N_{\rm eff}>3$, as in the case of neutrino asymmetries. 
On the other hand, electron neutrinos and antineutrinos 
are involved in the charged current weak processes which rule the neutron/proton chemical
equilibrium. Thus any change in the $\nu_e$ or $\bar{\nu}_e$ momentum distribution
can shift the neutron/proton ratio freeze out  temperature and in turn modify the
primordial $^4$He abundance. This is what happens for a non-zero $\nu_e-\bar{\nu}_e$
asymmetry at BBN, that shifts the neutron fraction towards
larger or smaller values for negative or positive values of $\eta_{\nu_e}$,
respectively.

We have performed an analysis of the effects of flavour neutrino asymmetries on the BBN outcome   in a similar way as in \cite{Mangano:2010ei}. We have solved the EOMs described in the previous section in a wide range of values for the total lepton (neutrino) asymmetry $\eta_\nu$, unchanged by oscillations, and the initial electron neutrino asymmetry $\eta_{\nu_e}^{\rm in}$. The obtained time-dependent neutrino distributions are then used as an input for the public numerical code $\mathtt{PArthENoPE}$ 
\cite{Pisanti:2007hk,parthenope}, as described in  \cite{Mangano:2010ei}. 
The computed abundances of both the ratio $^2$H/H and the $^4$He mass
fraction, $Y_p$, are compared with the
corresponding experimental determinations in order to find the allowed region of asymmetries 
$\eta_\nu$ and $\eta_{\nu_e}^{\rm in}$. Here, as in \cite{Mangano:2010ei}, we consider for 
the primordial $^2$H abundance the value 
\be
^2{\rm H/H}= (2.87 \pm 0.22) \times 10^{-5}\vv
\label{abundanceD}
\ee
obtained by averaging seven determinations from different 
Quasar Absorption Systems \cite{Iocco:2008va}. For the $^4$He mass fraction we 
use, as a conservative value, the result of the data collection analysis performed in
\cite{Iocco:2008va},
\be
Y_p=0.250 \pm 0.003\, . 
\label{he1}
\ee

\begin{figure}
\begin{center}
  \includegraphics[width=0.7\textwidth,angle=0]{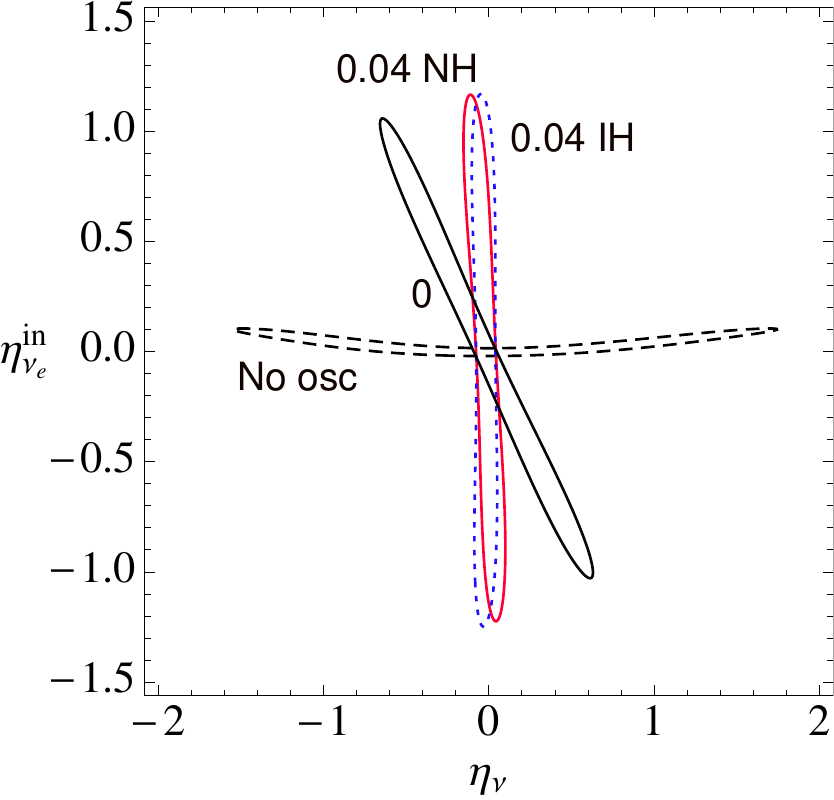}
   \end{center}
\caption{\label{bounds_eta} 95\% C.L.\ contours from our BBN analysis
in the $\eta_\nu-\eta_{\nu_e}^{\rm in}$ plane for several values of
$\sin^2\theta_{13}$: 0 (black solid line), 0.04 and NH (red solid line), 0.04
and IH (blue dotted line). The case of no neutrino flavour oscillations is shown for comparison as the black dashed contour.
}
\end{figure}

The main results of our BBN analysis are shown in 
figure \ref{bounds_eta} for the adopted determinations of $^2$H and
$^4$He. {}From this plot one can easily see the effect of flavour oscillations on
the BBN constraints on the total neutrino asymmetry. In the absence of neutrino 
mixing the value of $\eta_{\nu_e}$ is severely constrained by
$^4$He data, arising from a narrow region for the electron neutrino degeneracy,
$-0.018 \leq \xi_{e} \leq 0.008$ at 68\% C.L. Instead,
the asymmetry for other neutrino flavours could be much larger, since
the absolute value of total asymmetry is only restricted to the region 
$|\eta_\nu|\lesssim 2.6$ \cite{KangSteigman,Hansen:2001hi}. 
As we have previously seen, flavour oscillations modify this picture and an initially  large
$\eta_{\nu_e}^{\rm in}$  can be compensated by an asymmetry in the other flavours with opposite sign. The most restrictive BBN bound on $\eta_{\nu_e}$ applies then to the total asymmetry, an
effect that can be seen graphically in figure \ref{bounds_eta}  as a {\it rotation}
of the allowed region from a quasi-horizontal one for zero mixing to an almost vertical
region for $\sin^2\theta_{13}=0.04$, in particular for the IH.
In all cases depicted in figure  \ref{bounds_eta} the allowed values of the asymmetries are 
mainly fixed
by the  $^4$He bound, which imposes that the value of $\eta_{\nu_e}$ at BBN must be 
very close to zero,
while the data on primordial deuterium is crucial for closing the region.

\begin{figure}
	\begin{center}
     \includegraphics[width=0.85\textwidth,angle=0]{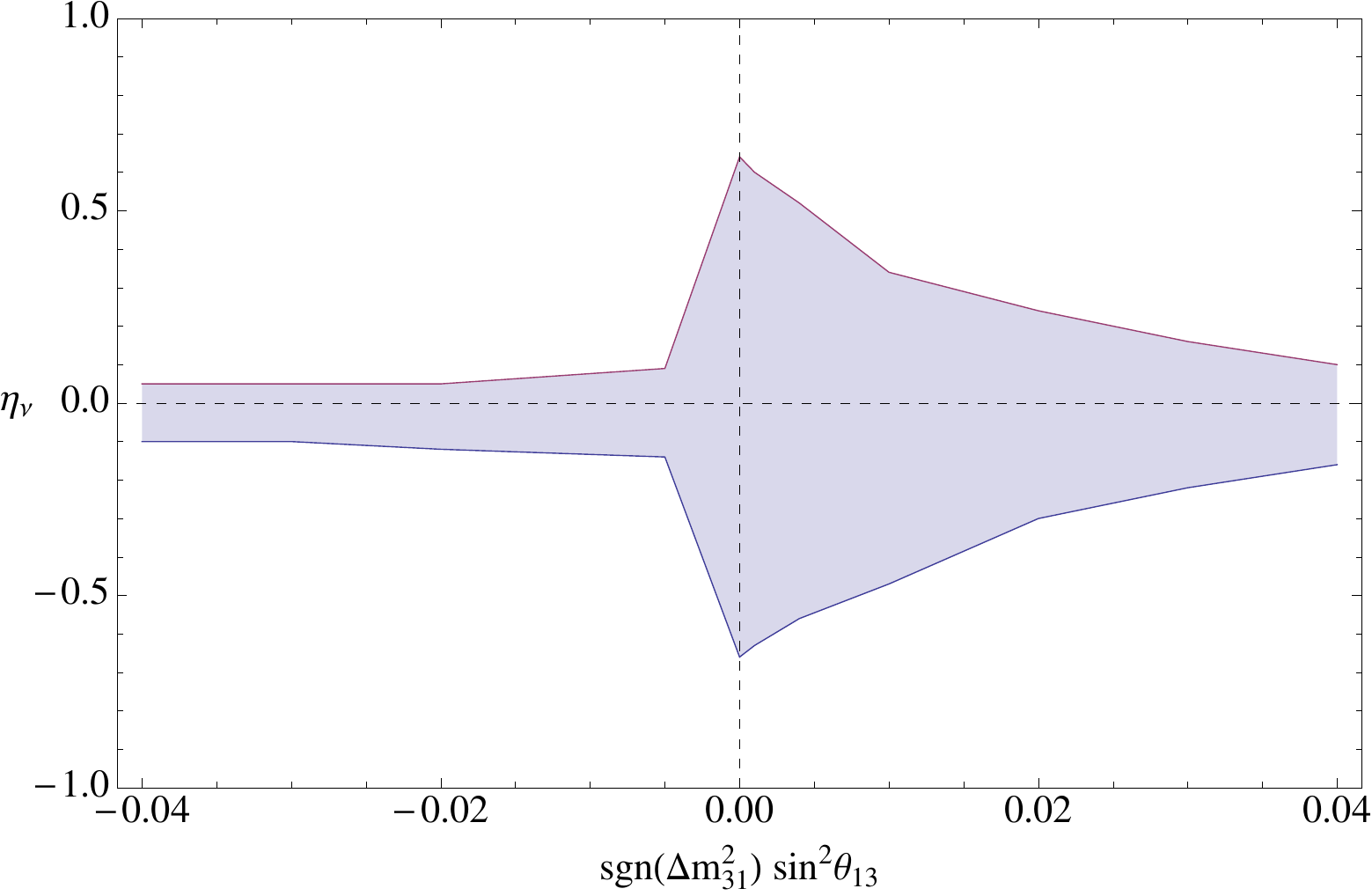}
     \end{center}\caption{\label{theta13_Lnu} The shadowed region corresponds to the
     values of the total neutrino asymmetry compatible with BBN at 95\% C.L., 
     as a function of $\theta_{13}$ and the neutrino mass hierarchy.}
\end{figure}

For values of $\theta_{13}$ close to the upper limits set by experimental data, the combined
effect of oscillations and collisions leads to an efficient mixing of all neutrino flavours before BBN. Therefore, the individual neutrino asymmetries have similar values, approximately
$\eta_{\nu_\alpha} \simeq \eta_\nu/3$, and the BBN bound on the electron neutrino asymmetry applies to all flavours, and in turn to $\eta_{\nu}$ as
considered in previous analyses
\cite{Iocco:2008va,
Barger:2003zg,Barger:2003rt,Cuoco:2003cu,Cyburt:2004yc,
Serpico:2005bc,Simha:2008mt}. We find that for $\sin^2\theta_{13}=0.04$
the allowed region at 95\% C.L.\ is 
$-0.17 (-0.1) \leq \eta_\nu\leq 0.1 (0.05)$ for neutrino masses following
a normal (inverted)  mass hierarchy. Note, however, that in the IH this result 
approximately holds for any value of $\sin^2\theta_{13}$ within the favoured region
by oscilation data, due to the resonant character of the conversions.
Instead, as discussed in \cite{Mangano:2010ei}, in the NH even values of order
$|\eta_\nu|\simeq 0.6$
are still compatible with BBN if $\theta_{13}$ is very small. The allowed
regions of the total neutrino asymmetry are depicted in figure \ref{theta13_Lnu} 
as a function of the mixing angle $\theta_{13}$ and the mass hierarchy.

\begin{figure}
	\begin{center}
     \includegraphics[width=0.85\textwidth,angle=0]{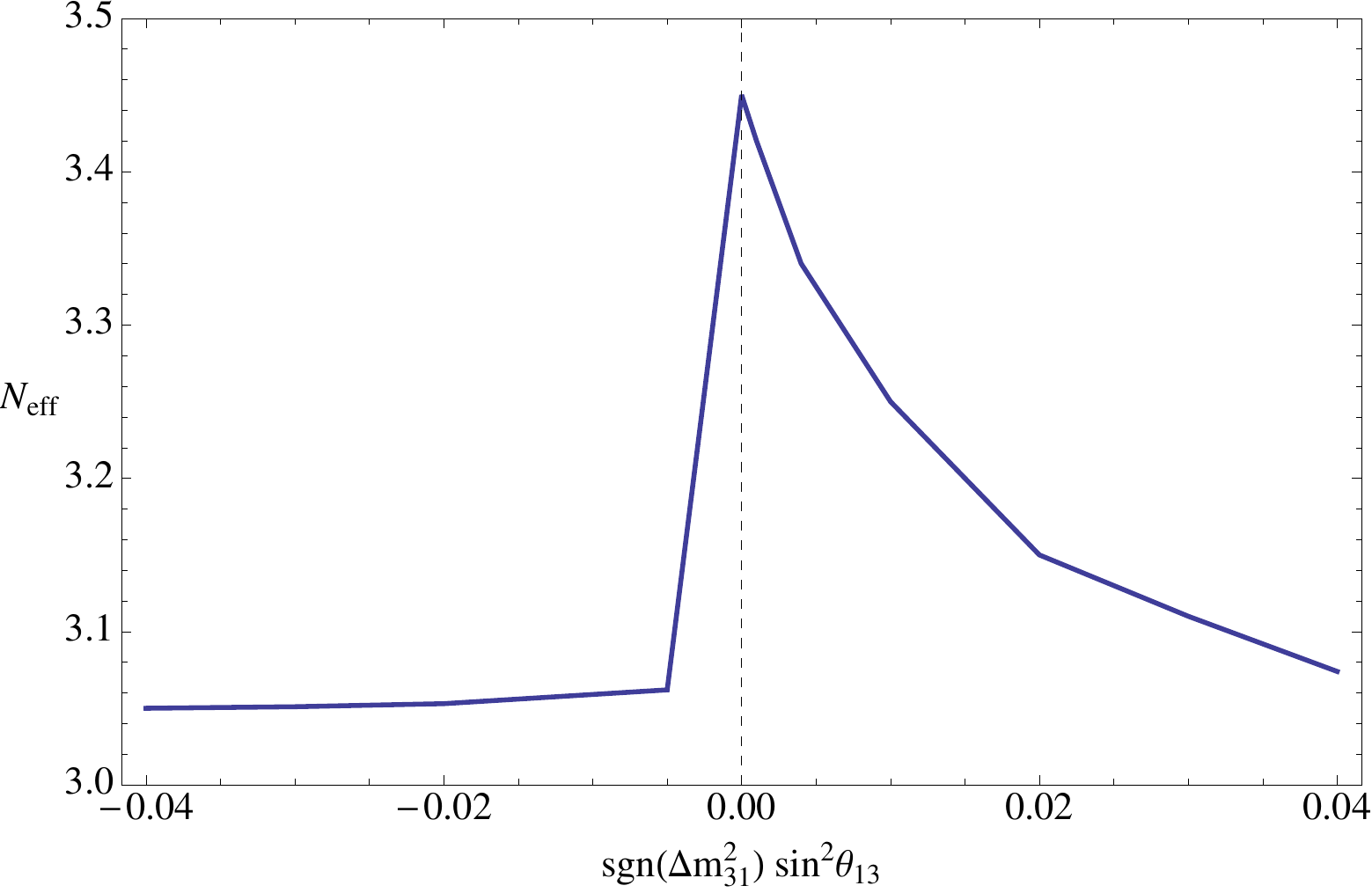}
     \end{center}
\caption{\label{theta13_Neff} Largest values of $N_{\rm{eff}}$ from primordial neutrino
asymmetries compatible with BBN at 95\% C.L., as a function of $\theta_{13}$
and the neutrino mass hierarchy.}
\end{figure}

Finally, the dependence of the largest value of $N_{\rm{eff}}$ from neutrino asymmetries
in the region compatible with BBN, as a function of the neutrino mass hierarchy
and the mixing angle $\theta_{13}$ is reported in figure \ref{theta13_Neff}. 
If the true value of $\theta_{13}$ lies in the upper part of the region favoured
by oscillation experiments (in particular T2K) or $\Delta m^2_{31}<0$,
the presence of primordial asymmetries can not lead to a contribution to the radiation energy density $N_{\rm{eff}}>3.1$\,.
On the other hand, for the NH and very small values of $\theta_{13}$, larger
values of $N_{\rm{eff}}$ are still compatible with BBN data, up to $3.43$ at 95\% C.L.

\section{Conclusions}
\label{sec:conclusions}

We have found the BBN constraints on the cosmological lepton number and its associated 
contribution to the radiation energy density, taking into account the effect of flavour neutrino oscillations. Once the other neutrino mixing parameters have been fixed by oscillation data, 
we have shown that pinpointing the value of $\theta_{13}$ is crucial to establish the degree of conversion of flavour neutrino asymmetries in the early Universe.

We conclude that the most stringent BBN bound on the total neutrino asymmetry, $|\eta_\nu|\lesssim 0.1$, requires that reactor neutrino experiments in the near future \cite{Huber:2004ug}, 
such as  Double Chooz, Daya Bay or Reno, confirm that the value of the third neutrino mixing 
angle is such that $\sin^2\theta_{13}\gtrsim 0.03$. This conclusion also applies to the whole allowed range of $\theta_{13}$ values at $3\sigma$ \cite{Fogli:2011qn,Schwetz:2011zk}
if neutrino masses follow an inverted hierarchy scheme. For smaller values of 
this mixing angle in the NH the BBN bound is relaxed, up to a maximum allowed range
of $-0.7\lesssim \eta_\nu \lesssim 0.6$ for $\theta_{13}=0$.

Similarly, a measured value $\sin^2\theta_{13}\gtrsim 0.03$ will imply that the maximum
contribution of neutrino asymmetries to the radiation content of the Universe can not exceed 
$N_{\rm eff}\simeq 3.1$, well below the expected sensitivity of the Planck satellite 
($0.4$ at $2\sigma$) \cite{Bowen:2001in,Bashinsky:2003tk,Hamann:2010pw}, whose first data release on the anisotropies of the cosmic microwave background (CMB) is expected in one year or so. 

Finally, we emphasize that BBN remains the best way to constrain a potential 
cosmological lepton number, despite the present precision of the measurements of the CMB anisotropies and other late cosmological observables. Bounds on the neutrino asymmetries
with these data \cite{Hamann:2007sb,Popa:2008tb,Shiraishi:2009fu,Castorina} do not improve 
those found in our work, but
are of course sensitive to other neutrino properties such as their masses.

\section*{Acknowledgements}

G.\ Mangano, G.\ Miele, O.\ Pisanti and S.\ Sarikas acknowledge support by
the {\it Istituto Nazionale di Fisica Nucleare} I.S. FA51 and the PRIN
2010 ``Fisica Astroparticellare: Neutrini ed Universo Primordiale'' of the
Italian {\it Ministero dell'Istruzione, Universit\`a e Ricerca}. S.\
Pastor was supported by the Spanish grants FPA2008-00319, 
FPA2011-22975
and Multidark CSD2009-00064
(MICINN) and PROMETEO/2009/091 (Generalitat Valenciana), and by the EC
contract UNILHC PITN-GA-2009-237920. This research was also supported by a
Spanish-Italian MICINN-INFN agreement, refs.\ ACI2009-1051 and AIC10-D-000543.






\begin{thebibliography}{00}


\bibitem{Schwetz:2011qt}
T.\ Schwetz, M.A.\ T\'ortola and J.W.F.\ Valle,
  New J.\ Phys.\  {\bf 13} (2011)  063004
  [arXiv:1103.0734].

\bibitem{Fogli:2011qn}
  G.L.\ Fogli, E.\ Lisi, A.\ Marrone, A.\ Palazzo and A.M.\ Rotunno,
Phys.\ Rev.\  D {\bf 84} (2011) 053007 [arXiv:1106.6028].

\bibitem{Schwetz:2011zk}
  T.\ Schwetz, M.A.\ T\'ortola, and J.W.F.\ Valle,
  New J.\ Phys.\  {\bf 13} (2011)  109401
 [arXiv:1108.1376].

\bibitem{Abe:2011sj}
  K.\ Abe et al.\ [T2K Collaboration],
  Phys.\ Rev.\ Lett.\  {\bf 107} (2011)  041801
  [arXiv:1106.2822].

\bibitem{Nunokawa:2007qh}
  H.\ Nunokawa, S.J.\ Parke, and J.W.F.\ Valle,
  Prog.\ Part.\ Nucl.\ Phys.\  {\bf 60} (2008)  338
  [arXiv:0710.0554].
  
\bibitem{Mangano:2005cc}
  G.\ Mangano, G.\ Miele, S.\ Pastor, T.\ Pinto, O.\ Pisanti and P.D.\ Serpico,
  Nucl.\ Phys.\  B {\bf 729} (2005) 221
  [arXiv:hep-ph/0506164].

\bibitem{Dolgov:2003sg}
  A.D.\ Dolgov and F.L.\ Villante,
  Nucl.\ Phys.\  B {\bf 679} (2004)  261
  [arXiv:hep-ph/0308083].
  
\bibitem{Kirilova:2006wh}
  D.P.\ Kirilova and M.P.\ Panayotova,
  JCAP {\bf 0612} (2006)  014
  [arXiv:astro-ph/0608103].

\bibitem{Melchiorri:2008gq}
  A.\ Melchiorri, O.\ Mena, S.\ Palomares-Ruiz, S.\ Pascoli, A.\ Slosar and M.\ Sorel,
   JCAP {\bf 0901} (2009) 036
  [arXiv:0810.5133].
  
\bibitem{Hamann:2011ge}
  J.\ Hamann, S.\ Hannestad, G.G.\ Raffelt, Y.Y.Y.\ Wong,
  JCAP {\bf 1109} (2011)  034 
  [arXiv:1108.4136].
  
  \bibitem{Komatsu:2010fb}
  E. Komatsu et al.,
 Astrophys.\ J.\ Suppl.\ Ser.\  {\bf 192} (2011) 18
 [arXiv:1001.4538].
  
\bibitem{MarchRussell:1999ig}
  J.\ March-Russell, H.\ Murayama, and A.\ Riotto,
  JHEP {\bf 9911} (1999)  015
  [hep-ph/9908396].

\bibitem{McDonald:1999in}
  J.\ McDonald,
  Phys.\ Rev.\ Lett.\  {\bf 84} (2000) 4798
  [hep-ph/9908300].


\bibitem{Schwarz:2009ii}
  D.J.\ Schwarz and M.\ Stuke,
  JCAP {\bf 0911} (2009)  025
  [arXiv:0906.3434].


\bibitem{Semikoz:2009ye}
  V.B.\ Semikoz, D.D.\ Sokoloff and J.W.F.\ Valle,
   Phys.\ Rev.\  D {\bf 80} (2009) 083510
  [arXiv:0905.3365].
 
\bibitem{KangSteigman}
  H.S.\ Kang and G.\ Steigman,
  Nucl.\ Phys.\  B {\bf 372} (1992) 494.

\bibitem{Hansen:2001hi}
  S.H.\ Hansen, G.\ Mangano, A.\ Melchiorri, G.\ Miele and O.\ Pisanti,
  Phys.\ Rev.\  D {\bf 65} (2002) 023511
  [arXiv:astro-ph/0105385].


\bibitem{Dolgov:2002ab}
  A.D.\ Dolgov, S.H.\ Hansen, S.\ Pastor, S.T.\ Petcov, G.G.\ Raffelt and D.V.\ Semikoz,
  Nucl.\ Phys.\  B {\bf 632} (2002) 363
  [arXiv:hep-ph/0201287].

\bibitem{Wong:2002fa}
  Y.Y.Y.\ Wong,
  Phys.\ Rev.\  D {\bf 66} (2002) 025015
  [arXiv:hep-ph/0203180].

\bibitem{Abazajian:2002qx}
  K.N.\ Abazajian, J.F.\ Beacom and N.F.\ Bell,
  Phys.\ Rev.\  D {\bf 66} (2002) 013008
  [arXiv:astro-ph/0203442].

  \bibitem{Lunardini:2000fy}
  C.\ Lunardini and A.Yu.\ Smirnov,
 Phys.\ Rev.\  D {\bf 64} (2001) 073006
  [arXiv:hep-ph/0012056].

\bibitem{Pastor:2008ti}
  S.\ Pastor, T.\ Pinto and G.G.\ Raffelt,
  Phys.\ Rev.\ Lett.\  {\bf 102} (2009) 241302
  [arXiv:0808.3137].

\bibitem{Mangano:2010ei}
  G.\ Mangano, G.\ Miele, S.\ Pastor, O.\ Pisanti and S.\ Sarikas,
  JCAP {\bf 1103} (2011) 035
  [arXiv:1011.0916].

\bibitem{Sigl:1993fn}
  G.\ Sigl and G.\ Raffelt,
  Nucl.\ Phys.\ B {\bf 406} (1993) 423.


\bibitem{McKellar:1994ja}
  B.H.\ McKellar and M.J.\ Thomson,
  Phys.\ Rev.\ D {\bf 49} (1994) 2710.

\bibitem{Pastor:2001iu}
  S.\ Pastor, G.G.\ Raffelt, and D.V.\ Semikoz,
  Phys.\ Rev.\  D {\bf 65} (2002)  053011
  [arXiv:hep-ph/0109035].
  
\bibitem{Notzold:1987ik}
  D.\ N\"otzold and G.\ Raffelt,
  Nucl.\ Phys.\  B {\bf 307} (1988)  924.
    
    
\bibitem{Pisanti:2007hk}
  O.\ Pisanti, A.\ Cirillo, S.\ Esposito, F.\ Iocco, G.\ Mangano, G.\ Miele and P.D.\ Serpico,
  Comput.\ Phys.\ Commun.\  {\bf 178} (2008) 956
  [arXiv:0705.0290].

\bibitem{parthenope}$\mathtt{PArthENoPE}$ web page:  http://parthenope.na.infn.it/

   \bibitem{Iocco:2008va}
  F.\ Iocco, G.\ Mangano, G.\ Miele, O.\ Pisanti and P.D.\ Serpico,
  Phys.\ Rep.\ {\bf 472} (2009) 1
  [arXiv:0809.0631].

\bibitem{Barger:2003zg}
  V.\ Barger, J.P.\ Kneller, H.S.\ Lee, D.\ Marfatia and G.\ Steigman,
  Phys.\ Lett.\  B {\bf 566} (2003) 8
  [arXiv:hep-ph/0305075].

\bibitem{Barger:2003rt}
  V.\ Barger, J.P.\ Kneller, P.\ Langacker, D.\ Marfatia and G.\ Steigman,
  Phys.\ Lett.\  B {\bf 569} (2003) 123
  [arXiv:hep-ph/0306061].

\bibitem{Cuoco:2003cu}
  A.\ Cuoco, F.\ Iocco, G.\ Mangano, G.\ Miele, O.\ Pisanti and P.D.\ Serpico,
  Int.\ J.\ Mod.\ Phys.\  A {\bf 19} (2004) 4431
  [arXiv:astro-ph/0307213].

\bibitem{Cyburt:2004yc}
  R.H.\ Cyburt, B.D.\ Fields, K.A.\ Olive and E.\ Skillman,
  Astropart.\ Phys.\ {\bf 23} (2005) 313
  [arXiv:astro-ph/0408033].

\bibitem{Serpico:2005bc}
  P.D.\ Serpico and G.G.\ Raffelt,
  Phys.\ Rev.\  D {\bf 71} (2005) 127301
  [arXiv:astro-ph/0506162].

\bibitem{Simha:2008mt}
  V.\ Simha and G.\ Steigman,
  JCAP {\bf 0808} (2008) 011
  [arXiv:0806.0179].

\bibitem{Huber:2004ug}
  P.\ Huber, M.\ Lindner, M.\ Rolinec, T.\ Schwetz and W.\ Winter,
  Phys.\ Rev.\  D {\bf 70} (2004) 073014
  [arXiv:hep-ph/0403068].

\bibitem{Bowen:2001in}
  R.\ Bowen, S.H.\ Hansen, A.\ Melchiorri, J.\ Silk and R.\ Trotta,
  Mon.\ Not.\ Roy.\ Astron.\ Soc.\  {\bf 334} (2002) 760
  [arXiv:astro-ph/0110636].

\bibitem{Bashinsky:2003tk}
  S.\ Bashinsky and U.\ Seljak,
  Phys.\ Rev.\  D {\bf 69} (2004) 083002
  [arXiv:astro-ph/0310198].

\bibitem{Hamann:2010pw}
  J.\ Hamann, S.\ Hannestad, J.\ Lesgourgues, C.\ Rampf and Y.Y.Y.\ Wong,
  JCAP {\bf 1007} (2010) 022
  [arXiv:1003.3999].
  
\bibitem{Hamann:2007sb}
  J.\ Hamann, J.\ Lesgourgues, and G.\ Mangano,
  JCAP {\bf 0803} (2008)  004
  [arXiv:0712.2826].
  
\bibitem{Popa:2008tb}
  L.A.\ Popa and A.\ Vasile,
  JCAP {\bf 0806} (2008)  028
  [arXiv:0804.2971].
  
\bibitem{Shiraishi:2009fu}
  M.\ Shiraishi, K.\ Ichikawa, K.\ Ichiki, N.\ Sugiyama and M.\ Yamaguchi,
  JCAP {\bf 0907} (2009)  005
  [arXiv:0904.4396].
  
 \bibitem{Castorina}
E.\ Castorina et al, in preparation

\end{thebibliography}



\end{document}